\documentclass[twocolumn]{aastex631}

\newcommand\irc{IRC$-$10412}

\shorttitle{A radio loud semiregular vairable}
\shortauthors{Luque-Escamilla \& Mart\'{\i}}
\graphicspath{{./}{figures/}}

\begin{document}

\title{A radio loud semiregular variable}

\author[0000-0002-3306-9456]{Pedro L. Luque-Escamilla}
\affiliation{Departamento de Ingenier\'{\i}a Mec\'anica y Minera, \\
Escuela Polit\'ecnica Superior de Ja\'en, Universidad de Ja\'en,\\
Campus Las Lagunillas s/n, 23071 Ja\'en, Spain}

\author[0000-0001-5302-0660]{Josep Mart\'{\i}}
\affiliation{Departamento de F\'{\i}sica,\\
Escuela Polit\'ecnica Superior de Ja\'en, Universidad de Ja\'en,\\
Campus Las Lagunillas s/n, 23071 Ja\'en, Spain}



\begin{abstract}
As a byproduct of our search for Galactic stellar systems with gamma-ray emission, we have identified an unrelated cool and evolved star (\irc) 
that attracted our attention due to its strong radio emission level with a spectral index matching, almost perfectly, the canonical +0.6 value expected
from an ionized stellar wind. A follow-up observational analysis was undertaken given that these two properties are hard to reconcile as
originating in the same stellar object. As a result,  \irc\ has been classified as a new semiregular variable of SRb type
in the asymptotic giant branch, and different but consistent 
estimates of its mass-loss parameter are reported. We  propose that its
unusually high radio emission arises from a $\sim 10^{-5} M_\odot$ yr$^{-1}$  stellar wind exposed to an external source of ionizing photons, possibly
coming from nearby OB associations. 

\end{abstract}



\section{Introduction} \label{sec:intro}

Mining the multi-wavelength data archives currently available provides ample room for discovery and scientific exploitation.
Moreover,  research in a given area of Astrophysics sometimes leads to unexpected findings in another. This is the case reported here.
In the course of our efforts to identify compact radio sources of early-type stellar nature that
could be behind potential gamma-ray emitters \citep[e.g.][]{2020MNRAS.492.4291M, 2021Univ....7..214M, 2023MNRAS.518.3017L},
we came across an interesting object worth of being reported for its scientific interest
in the domain of cool, evolved stars. 
These objects at the end of their lives often behave as long-period variables (LPVs),
which are
specially interesting 
as independent distance calibrators via period–luminosity relations \citep{2014White, 2019Rau, 2021Trab}. 
Moreover, these aged stars are major contributors to the processed material in the interstellar medium \citep[\textit{e.g.}][]{Sedlemayr94, Karakas_Lattanzio_2014}. 
Estimating their mass-loss rates is therefore of great astrophysical interest 
\citep{2000Winters}.

Stars at such advanced stages of evolution
 are also known to be sources of thermal radio emission originating in their
winds and envelopes. Our understanding of this phenomenon is mostly based around the concept of stellar radio photosphere. Its 
radius corresponds to optical depth unity of the dominant free-free emission mechanism, 
that usually amounts to a few times the optical photospheric radius
\citep{1997ApJ...476..327R, 2018AJ....156...15M}. The observed radio brightness for giant and supergiant stars closer than
1 kpc  are typically below the 1 mJy level \citep{1991AJ....101..230D, 2007astro.ph..3669R}.
Radio studies of these systems and their stellar activity have seen their motivation enhanced in recent times
given their connection with exo-space weather phenomena and life sustainability in exo-planets 
\citep{2018ASPC..517..369C}.

In this work, we report the identification of a new asymptotic giant branch (AGB) LPV star with heavy absorption from a circumstellar dust that exhibits a remarkable differential feature:
an intense radio emission exceeding by at least two orders of magnitude
the typical radio luminosity of other members of its class.
This unique behavior opens the possibility to an independent  estimation of its mass-loss rate with no need of empirical prescriptions, together with engrossing the so-called dust-enshrouded AGB star family.

The rest of the manuscript is organized as follows: the next section introduces the target 
of our study, the system \irc, and how we focused our attention on it.  The paper continues with  a section devoted to present the multi-wavelength
observational evidence followed by another section of discussion  and modeling of the spectral energy distribution (SED). 
Conclusions are summarized at the end.

\section{The elusive nature of \irc}

\irc\ (a.k.a. IRAS 18179$-$1346 and StRS177) is one of entries in the historical Infrared Catalog (IRC) compiled by \citet{1969NASSP3047.....N}.
After a first unsuccessful classification attempt  \citep{1973AJ.....78..669H}, the star  was photometrically  observed by
\cite{1985ApJS...58..167L} who assigned it a supergiant M7.0 spectral type. In addition, these authors provided the first suspicion of \irc\ variability as they consider it a possible member of the so called "M10 Mira variables".
{\it IRAS} low-resolution spectra, reported by \citet{1989AJ.....98..931V} and later \cite{1997ApJS..112..557K}, displayed a 
featureless continuum consistent with a M star. This interpretation was at some point challenged given the apparent
non-detection of TiO bands in the photographic-infrared objective prism survey by \citet{1992AJ....103..263S}.
Together with strong H$\alpha$ emission detected in the optical by \cite{2000ApJS..131..531R}, these two last works
pointed to a highly-reddened, early-type classification instead of a late evolved star. 
Independently, the Very Large Array (VLA) survey of the Galactic Plane, conducted by \citet{1994ApJS...91..347B},
revealed the radio source GPS5 017.223+0.395, whose position is very close to that of \irc, and 
then believed to be an ultra compact HII region (UCHII).
 It was studied as such in follow-up work,
 but this GPS5 object was noticed not to agree well with the typical properties of high-mass star formation sites,
 in particular its non-flat spectral index \citep{2009A&A...507.1467L}.
 A possible UCHII mis-classification was even suspected  
\citep{2009ApJS..181..255A}. In recent times, the {\it Gaia} Data Release 3 \citep{2023A&A...674A..15L}
 has provided the latest improvement of our knowledge about
\irc\ with a trigonometric parallax distance of $d = 1.6 \pm 0.1$ kpc, which will be adopted here. 
They also find additional hints of being a LPV
around a mean apparent magnitude G=12.48. 

In parallel, \irc\ would not have attracted the attention of the authors if it had not been for our attempt to
search for possible stellar gamma-ray sources among highly reddened luminous stars (to be reported elsewhere). 
Indeed, a significant fraction of currently known gamma-ray binaries have luminous, early-type optical counterparts
\citep{2019RLSFN..30S.107P}, in addition to compact radio emission. Why not similar obscured systems could be waiting for discovery?
Our starting point was the \citet{1992AJ....103..263S} list of highly reddened objects with unbanded spectra
suspected to be distant luminous stars. We cross-identified them with different radio surveys, in particular the
NRAO VLA Sky Survey (NVSS) by \citet{1998AJ....115.1693C}. At this point, \irc\ emerged as one of the matches
deserving additional study after exploring the literature quoted in the previous paragraph.
We also noticed that \irc\  had been detected, and  proposed to be a likely radio-star, in the 5 GHz CORNISH survey 
\citep{2012PASP..124..939H}, hence encouraging further investigation.

\section{Observational properties}

\subsection{The radio spectrum and morphology}.   \label{radiopart}

\begin{table}
\caption{Radio flux densities of \irc}             
\label{radiosp}      
\centering                          
\begin{tabular}{c c c}        
\hline\hline                 
Freq.  &           Integrated                &    \\
(GHz)      &        Flux Density              &     Reference                \\
               &             (mJy)                     &                    \\
\hline                        
 1.4            &   $87.9 \pm  2.7$               &      NVSS, \cite{1998AJ....115.1693C}      \\
 \hline
 1.4            &    $99^{*}$                               &       GPS5, \cite{1994ApJS...91..347B}      \\   
 \hline
 3.0            &  $157.43 \pm  0.56$          &      VLASS1 and VLASS2$^{**}$                      \\
                  &  $160.29 \pm  0.56$          &       {\small https://science.nrao.edu/vlass}  \\
\hline
 5.0            &  $199.86 \pm 19.69$         &     CORNISH, \cite{2012PASP..124..939H}  \\  
 \hline
 5.0            &   $151.6^{*}$                           &      GPS5, \cite{1994ApJS...91..347B}   \\      
 \hline
42.95        &  $595 \pm  90$                 &    \cite{2012PASP..124..939H}    \\
 \hline                                   
\end{tabular}
\\(*) {\small Error unavailable.} (**) {\small Measured using AIPS JMFIT.}
\end{table}

\begin{figure}
\epsscale{1.30}
\plotone{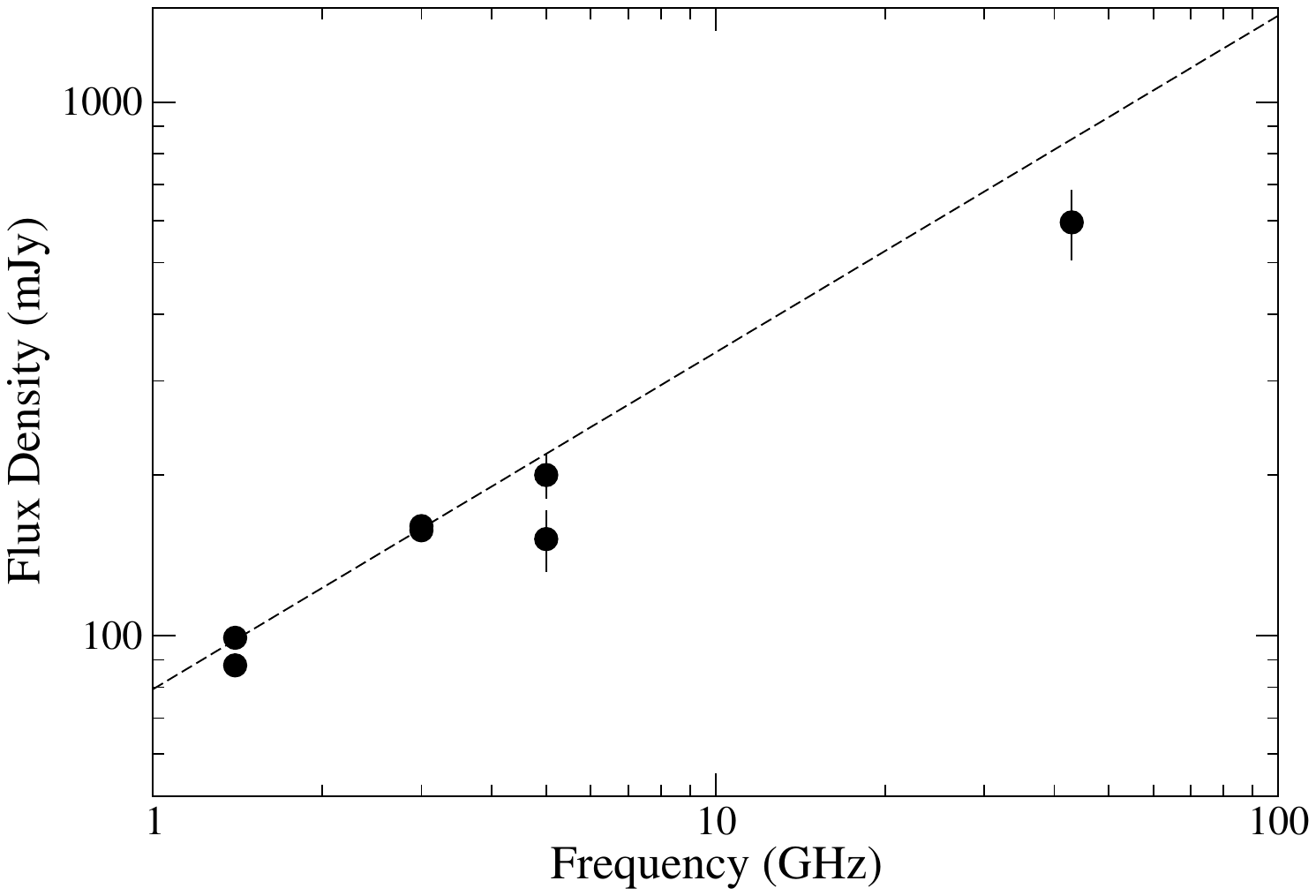}
\caption{Radio spectrum of \irc, in the cm-mm wavelength domain, assembled using the  survey data 
listed in Table \ref{radiosp}. The dashed line is a simple power-law fit.
\label{sp}}
\end{figure}

\begin{figure}
\epsscale{1.0}
\plotone{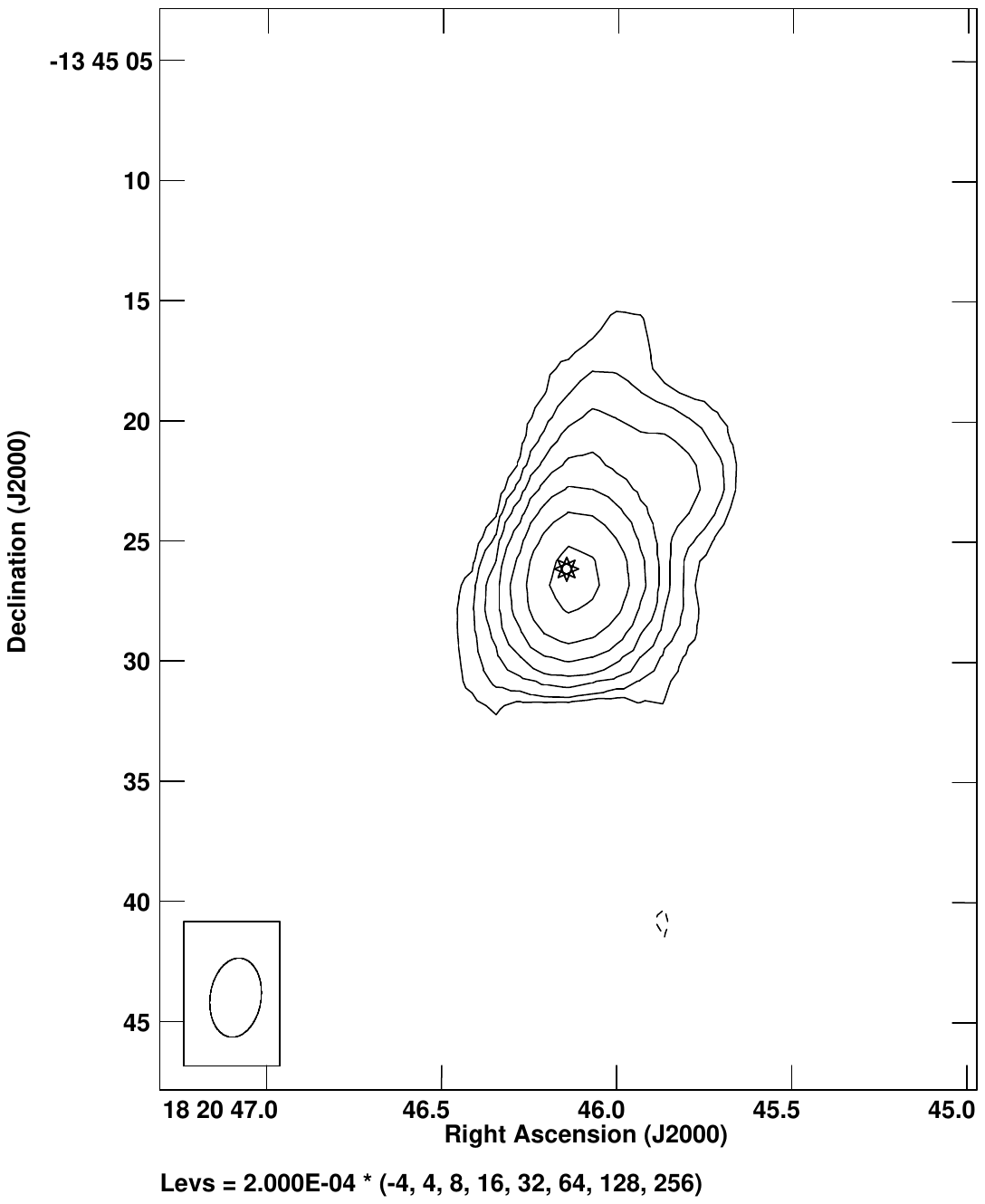}
\caption{VLSS1 resolved radio map of \irc\ at the 3.0 GHz frequency with observation date 2019 June $19^{th}$. 
The bottom left ellipse represents the clean synthesized beam of $3.31 \times  2.12$ arc-second$^2$, with position angle $-8.0^{\circ}$. The star symbol indicates the {\it Gaia} DR3 optical position of \irc.
\label{vlass}}
\end{figure}

Radio flux densities of \irc\ can be found in the literature from different sources. We have collected them in
Table \ref{radiosp}. In addition, 
we have also used the recent VLA Sky Survey (VLASS) 
maps\footnote{\tt https://science.nrao.edu/vlass}, and directly measured the target flux density using the JMFIT task of the
AIPS software package for the two epochs available. The result is plotted in Figure \ref{sp}. 
Neglecting radio variability, a simple least squares fit indicates that the
radio spectrum is well represented by $S_{\nu} = (79 \pm 4)~{\rm mJy} \left[ \nu/{\rm GHz}   \right]^{+0.63 \pm 0.05}$,
where $\nu$ is the observing frequency.
On the other hand, the VLASS angular resolution is high enough to resolve the \irc\ morphology at 3.0 GHz  as evidenced in the
contour map of Figure \ref{vlass}. By fitting an elliptical Gaussian using JMFIT, the deconvolved angular size turns out to be
$(3.5 \pm 0.1)\times(2.87\pm0.04)$ arc-second$^2$, with position angle $165^{\circ} \pm 3^{\circ}$.



\subsection{The \irc\  optical light curve}  	\label{lctone}

\begin{figure}
 	\plotone{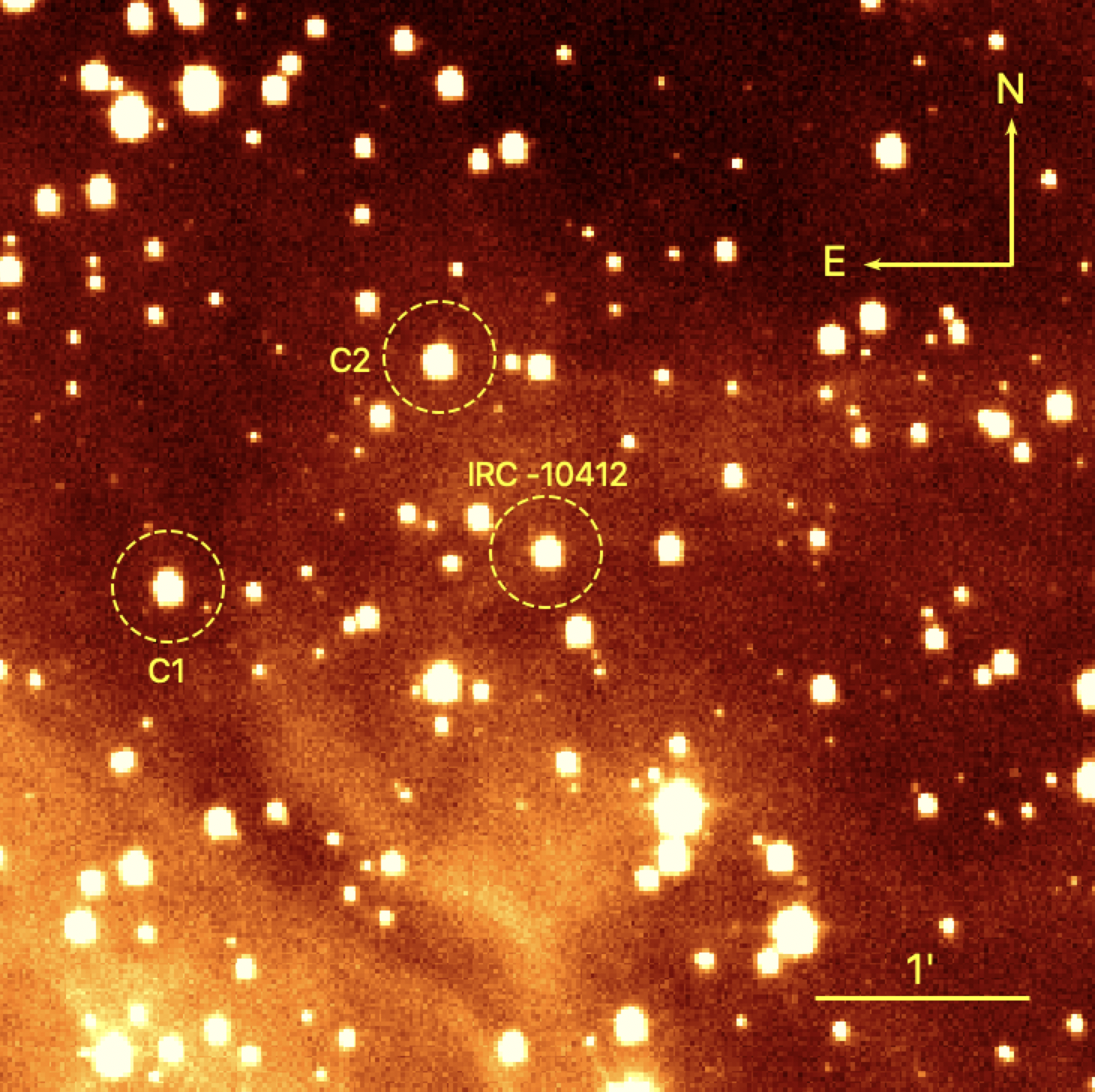}
 	\caption{ZTF $r$-band image of  the \irc\ field. The orientation and image scale are given by the yellow bars. The target and comparison stars are
 		marked using dashed circles.
 		\label{zr_image}}
 \end{figure}

Given the hints of optical variability mentioned above, we decided to further explore this issue. Although \irc\ is much brighter in the infrared,
it is still accessible in red optical filters where most dedicated variability surveys are at work. In particular,
 the Zwicky Transient Facility (ZTF)  \citep{2019PASP..131a8003M} appeared as the most suitable for this purpose with hundreds
 of images of available for differential photometry. A stacked ZTF  view of the \irc\ field of view in its $r$-band is shown in Figure \ref{zr_image}
 displaying both the central target and our two comparison stars labelled as C$_1$ and C$_2$.
 
\begin{figure}
\epsscale{1.35}
\plotone{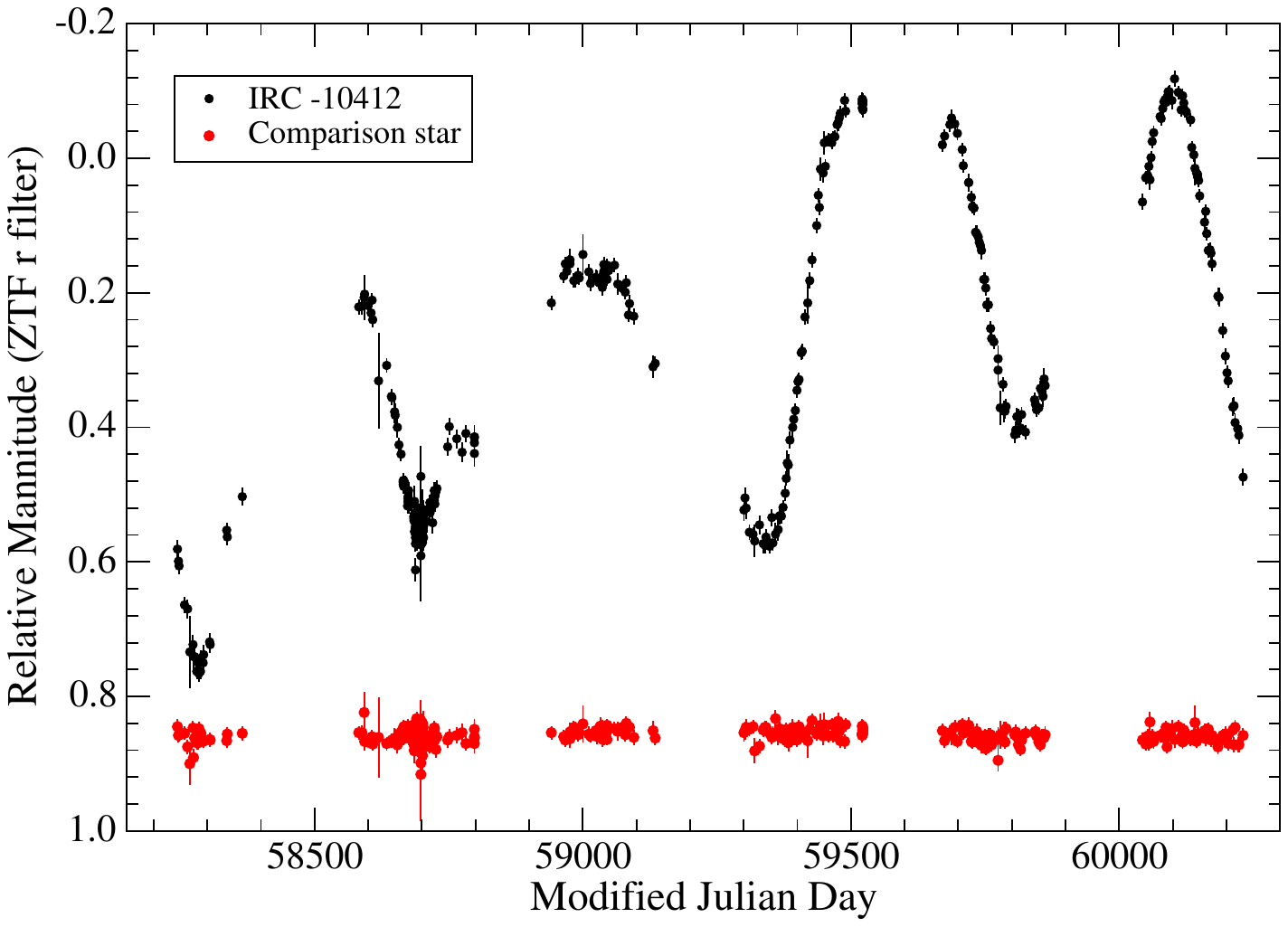}
\caption{Photometric analysis of \irc\ using  ZTF $r$-band images. Black points correspond to the target light curve.
Red points represent the difference between comparison star C$_1$ and the check star C$_2$, that remains stable within about 0.02 magnitudes.
\label{lc}}
\end{figure}

 These two objects correspond to  stars Gaia DR3 4152417198316110336 and 
Gaia DR3 4152420221973094144, respectively. We downloaded the ZTF frames and conducted our own differential photometry on them
using the IRAF software package. The resulting light curve is plotted in Figure \ref{lc} where strong variability with amplitude of
$\sim 1$ magnitude on time scales of hundreds of days is evident. When comparing with the light curve
produced by the ZTF pipeline, whose mean level is $r=15.30$ magnitudes, the same variability behavior is obtained .

\begin{figure}
\epsscale{1.25}
\plotone{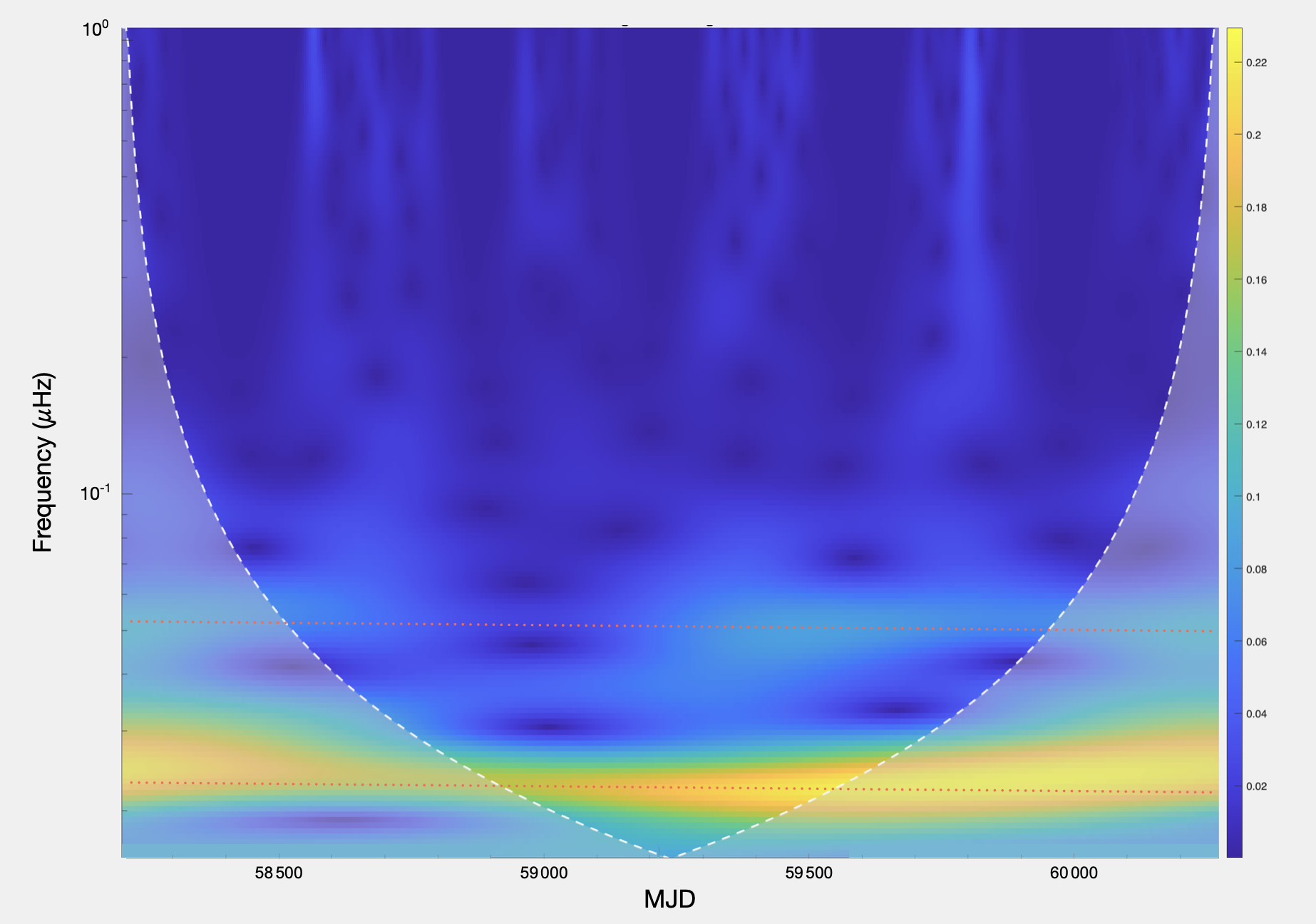}
\caption{Wavelet periodogram of the ZTF $r$-band light curve of \irc. Two main periods of 553 and 219 days are shown with dashed horizontal lines.
\label{wavelet}}
\end{figure}

To better assess the variability time scale we carried out a period search using different methods. 
Phase Dispersion Minimization \citep[PDM;][]{1978PDM} Lomb-Scargle \citep{Scargle_2002} and wavelet analysis (see Figure \ref{wavelet}) yield the same result: the star light curve appears to evolve under two main periods, estimated as 219 and 553 day, that stand out above the rest of spectral features.
Although the 219 days period could be the one year alias of the fundamental mode, the CLEAN algorithm, which deals better with sampling problems, clearly maintains the two mentioned periods.





\subsection{Assembling the \irc\ SED}

\begin{figure}
\epsscale{1.20}
\plotone{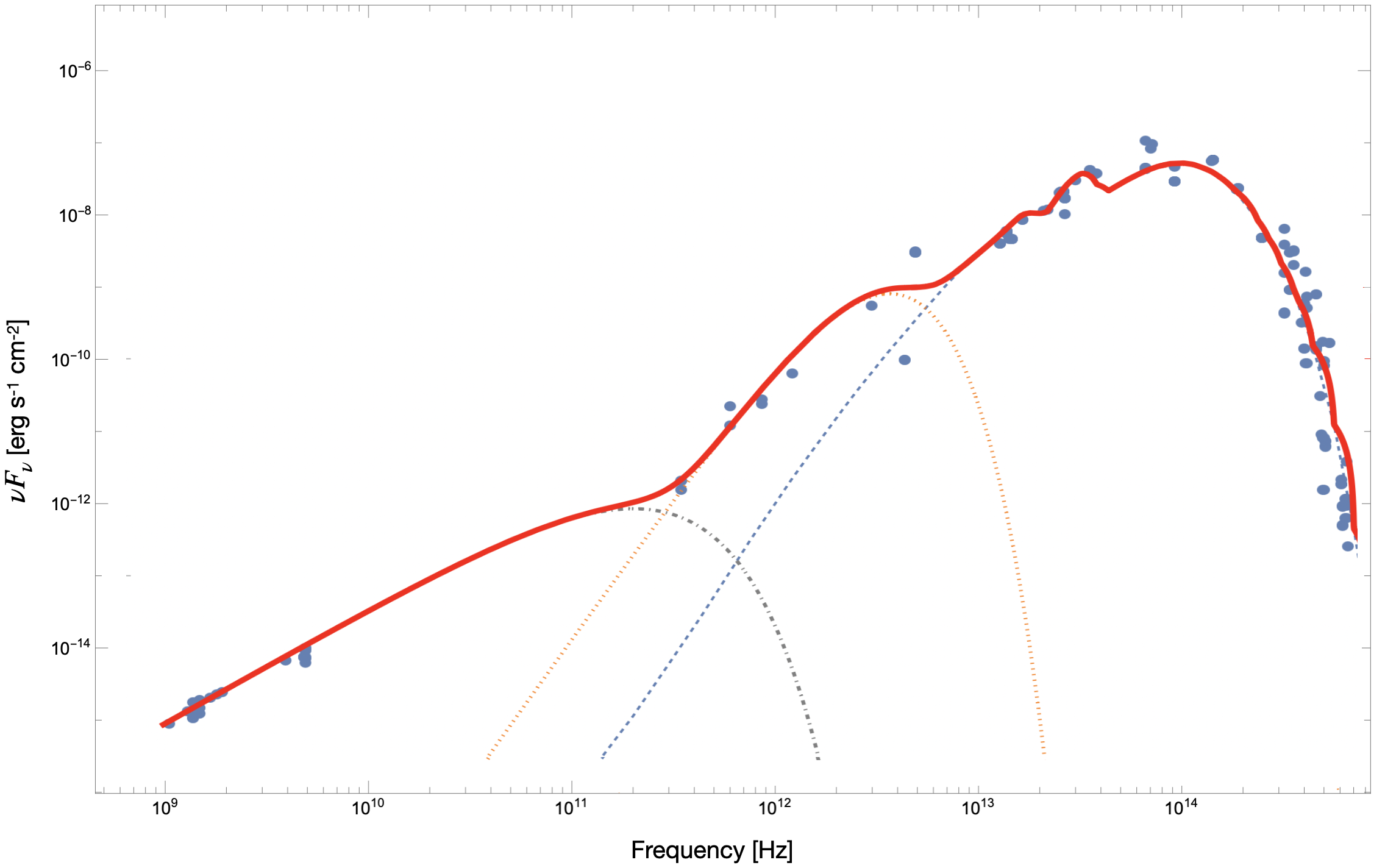}
\caption{Dereddened SED of \irc\ and model fit given by the continuous red line. How the fit has been assembled in the different
wavelength regions is explained in the text. The dotted lines with black, orange and blue colors correspond to the components modeled as
ionized wind radio emission, external cold dust  emission and inner dust shell emission, respectively.
\label{sed}}
\end{figure}

In order to facilitate the discussion of the physical nature of \irc, we took advantage of the 
spectral energy distribution (SED) building tools provided by 
the Strasbourg astronomical Data Center\footnote{{\tt https://vizier.cds.unistra.fr}, \cite{2000A&AS..143...23O} and references therein.}. 
We have built an empirical broad-band SED by using the multi-wavelength data available from 
2MASS, 
WISE,
Palomar POSS/ZTF, 
\textit{Gaia}, 
SDSS, 
Hipparcos, 
IRAS, 
AKARI, 
UKIDS/UKIRT, 
Pan-STARRS, 
SkyMapper, 
MSX, 
together with the radio data collected in Subsection 3.1.
Optical and infrared fluxes have been dereddened from interstellar extinction taking into account the
model of \citet{2006Marshall} and the \textit{stilism} app based on the work of \citet{2017Capitanio}.
As a result, we estimate the interstellar absorption as $A_V = 3.1 \pm 0.6$, assuming an optical parameter $R_V = 3.1$. The corresponding
absorption coefficients for other bands were obtained from the SVO profile filter \citep{2020SVO}. The final plot is shown in Figure \ref{sed} as a function of frequency. 


\section{Results and discussion}

The \textit{Gaia} DR3 release delivers for our source 
 $T_\mathrm{eff} = 3722\pm4$ K, $\log (g/\rm{cm~ s}^{-2}) = 0.125^{+0.012}_{-0.007}$  and a chemical abundance [M/H]$=-0.82$ dex
 computed with its Apsis pipeline based on low-resolution BP/RP spectra (GSP-Phot module).
These values are
 typical for cool, giant, evolved stars.
Moreover, they are consistent with the M spectral type proposed in the literature for \irc .
However, the parameters from Apsis for this kind of stars do not
consider the effects of circumstellar envelopes and compositional changes due to dredge-up.
In fact, \citet{2023Messineo} shows that while gravity measures are not far from literature values, $T_\mathrm{eff}$ is overestimated and [M/H] is poorly constrained.
In any case, chemical abundance in \irc\ should not exceed $0.5$ dex, and its $T_\mathrm{eff}$ must be close to 3000 K, in accordance with its suspected cool, evolved stellar nature. 

\subsection{Classifying \irc} \label{clasi}

\irc\ is clearly an optical variable source on time scales of hundreds of days thus confirming
the {\it Gaia} LPV classification. These objects are commonly related to cool stars in the last stages of their evolution, 
mainly Miras, AGB stars (with masses below about $8 M_\odot$), and the more massive red supergiants (RSGs).
The main cause under this variability seem to be pulsation. 
Although the driving mechanism is not well known 
(see \citet{2007uvs..book.....P} for an introductory review), 
much effort have been recently devoted to obtain a improved modeling \citep[see, for instance,][]{2019Trabucchi, 2021Trabucchi}. 

The variability cycles are often related to the fundamental and higher overtone pulsations.
With al caution, it is tempting to associate the light  curve 553 days periods found in section \ref{lctone} with the  fundamental mode. If the 219 days period was real, it could be related to the first overtone. 
On the other hand, the observed optical variability is small compared to the expected in Mira systems \citep[$\gtrsim2.5$ mag in $V$ filter, moving down to $\gtrsim 0.4$ mag for $K$ filter, \textit{e.g.}][]{1994Whitelock}.
Taking all these facts, \irc\ qualifies to be considered a likely semi-regular variable SRb.

As AGBs show larger amplitudes in their pulsations than RSGs, the  $\sim 1$ mag bright variations of \irc\ seem to be  more suitable to a star of the first kind.
Here we are assuming that the observed peak-to-peak amplitude  is intrinsic to \irc.
A non-variable infrared component, either internal to the \irc\ system or unnoticed inside the photometric aperture, could be reducing the true variability amplitude. However, we do not have evidence that this is actually one of these cases.

 Both AGBs and RSGs are highly reddened stars, probably associated with the absorption of their light output by dust grains and its posterior re-emission at longer wavelengths.
Significant extinction 
in the optical together with  excess emission in the
thermal infrared is then a clear indication of this scenario. This seems to be the case in the 
\irc\ SED  displayed in Figure \ref{sed}. 
In order to get a deeper insight into the properties of our source,
we will proceed trying to fit the observed broad-band SED regions having in mind that the available data points are not strictly simultaneous in time.


	\subsection{Radio counterpart association with \irc}

The 1-$\sigma$ uncertainty for the the CORNISH radio survey,
where the best radio position comes from at the present time, amounts to 0.147 arc-second.
The coincidence between \irc\ and the proposed radio counterpart is at the 2.6 and 1.4-$\sigma$ level
in right ascension and declination, respectively.  This is a pretty
good agreement for the radio photocenter and the optical {\it Gaia} DR3 position considering that we are dealing with a slightly extended radio source.
Given this small offset, we consider that the chances of association are very high and will proceed under the assumption that radio and optical/IR emission come from the same object. 
Moreover, an extragalactic source mimicking so well the radio spectrum and extended morphology of a stellar wind will be very unlikely.

\subsection{SED radio modelling and mass-loss estimate}

A specially striking characteristic of our source is its unusually high level of radio emission with respect to typical cool, evolved stars. For instance, the RSG 
VY CMa integrated flux density ranges from $0.24 \pm 0.02$ mJy at $8.4$ GHz to  
$2.5 \pm 0.5$ mJy at $42.95$ GHz \citep{2005Lipscy}, which is at least two orders of
magnitude smaller than in our source. This contrast cannot be accounted for the difference in distance and is even more dramatic for other LPVs
\citep{1996Reid}.
In addition, these authors find that the power-law spectral index 
in these objects is of the order of $1.7$-$2$, pointing to an optically-thick
emission that they modeled as a ``radio photosphere''.
However, this value is far from the spectral index of $0.6$ observed for \irc\ (see  \ref{radiopart}),
Another possible scenario involves an asymmetric, ionized region produced when the stellar wind of our cool, evolved star  is ionized by a hot companion, like a white dwarf.  The strong H$\alpha$ line emission reported by  \citet{2000ApJS..131..531R} would fit with this picture, that resembles a symbiotic Mira \citep{1984STB,1984STB2,1994Seaquist}. 
The well-known STB model   proposed by these authors is capable to explain radio emission based on a single ionization parameter $X$ which defines the geometry of the ionized region, and depends on the red giant mass-loss rate, the binary separation, and the Lyman continuum luminosity of the ionizing source.  If we assume reasonable typical values for the sum of masses of cool giant and white dwarf from $2$ to $10$ $M_\odot$ and a period of about 2 years, we get a binary separation $a \sim 10^{13}$ cm. 
Taking into account the mass-loss rate and wind terminal velocity derived below from infrared data,
we finally obtain an ionization parameter $X \sim 10^{-4}$. However, $X > \pi/4$ is needed to get the radio spectral index of $0.6$ observed for \irc\ \citep{1984STB2}. This fact, together with the lack of any conspicuous period of the order of years in the light curve, led us to discard this binary scenario.

Therefore, we finally propose that radio emission comes from a spherically symmetric stellar wind whose density decays as the inverse of the distance squared \citep{1975A&A....39....1P,  1975MNRAS.170...41W}, which theoretically gives an spectral index of $0.6$, remarkably coincident with that observed for \irc. Thus, we would be facing a
thermal free-free emission in a completely ionized outflow from the star, in spite of its intrinsic cool nature. Of course, an external source of ionizing photons needs to be invoked if this is the case.
These facts could be reconciled if we take into account the close  vicinity of the Serpens 2
and Scutum 3 OB associations \citep{1995AstL...21...10M}. Under this assumption,
we fitted the observed SED at radio wavelengths with the  \citet{1975A&A....39....1P} model together with an exponential cutoff (at $\nu_c = 10^{11}$ Hz), 
adopting a plasma electron temperature of  $T_e=10^4$ K for a completely ionized wind, an average ionic charge  $\bar{Z} = 0.9$ for a cold star with neutral Helium, and a typical mean atomic weight per electron $\mu=1.2$.
We have also fixed a  wind terminal velocity of $17.5$ km s$^{-1}$, which is a mean value between the  $10$ to $25$ km s$^{-1}$ typical range for AGBs and RSGs, and is appropriate for nearby dust-enshrouded AGBs \citep{2001OLIVIER}. 
In this way, 
we get the best fit for a mass-loss rate of $\dot{M} = 1.05\times 10^{-5} M_\odot$ yr$^{-1}$  (dot-dashed black line in Fig. \ref{sed}).
It is noteworthy that the  \citet{1975A&A....39....1P} formalism leads us to directly obtain the mass-loss rate without the need of an optical-infrared SED fit or a empirical prescription, as usual for cool, evolved stars. This historic model also allows us to estimate the angular diameter of the radio emitting region,
 yielding $\sim 1.8^{\prime\prime}$. It is really noteworthy that this value compares well, within a factor of two, with the deconvolved
angular size obtained in subsection \ref{radiopart}
from the VLSS map in Figure \ref{vlass} ($\nu=3.0$ GHz).
Despite the model simplicity, this close coincidence is reassuring with respect
to the correctness of the stellar wind interpretation.

%
%
%

\subsection{SED optical-infrared modelling and envelope parameters}\label{sedfit}

The \irc\ SED shows a clear infrared excess and optical absorption as previously stated, 
which lead us to suspect the existence of a dense circumstellar envelope. We have then modeled the optical-infrared emission  with the code DUSTY \citep{DUSTY}, which solves the radiative transfer for a spherically symmetric dust distribution around a central source. For the theoretical fit shown in Figure \ref{sed} 
(blue dashed line),
we need to set an opaque optical depth $\tau_V = 10.8$  at $0.55$ $\mu$m. The central star is modeled with a blackbody with $T_\mathrm{eff} = 3300$ K.  
For the chemical composition of the dust we take into account the existence of an observed peak around $10\, \mu$m which is characteristic
of silicate grains, although our best result is obtained with a mixture of 40\% of astronomical silicate and a 60\% of graphite with properties adopted from \citet{Draine}. 
We also used a standard MRN grain size distribution \citep[a power-law with exponent $-3.5$;][]{1977MATHIS} typical of cool, evolved stars \citep[e.g.][]{vanLoon1}.
To derive the mass-loss rate one has to know the dust grain density and the gas-to-dust mass ratio $r_{gd}$. 
For the former we consider a default $\rho_d = 3$ g cm$^{-3}$, but the ratio $r_{gd}$ is itself not well constrained. Different studies for classical RSGs allow a comparison between gas mass-loss rates \citep[\textit{i.e.}][]{debeck2010} and dust mass-loss rates \citep[\textit{i.e.}][]{Verho2009} returning ratios between tenths to several hundreds. 
Similar values have been measured for AGBs \citep{1985Knapp, 2018Dhar}.
Here, we have adopted an intermediate  $r_{gd} = 350$, which is the expected value for the galactic longitude of \irc\ \citep{2002Groen}. 
The density distribution is based upon a hydrodynamic computation of a dust-driven wind at constant mass-loss rate, with an asymptotic dependence with the distance from the star $\rho \sim r^{-2}$.
The innermost part of the dust shell in our fit is at a temperature $T_{in} = 1500$ K, which is a  commonly adopted value for 
dust sublimation. 
The shell is assumed to extend up to $10^4$ times its inner radius (so we define $\xi \equiv R_{out}/R_{in}= 10^4$), such that the dust density is low enough at the outskirts  so that it has no effect on the spectrum.
The DUSTY parameters corresponding to the blue dotted line fit displayed in  Figure \ref{sed} are collected in Table \ref{tablaDUSTY}.

\begin{table}
	\centering
	\caption{Output of DUSTY model used.}
	\label{tablaDUSTY}
	\begin{tabular}{lr} 
		\hline
		Parameter & Value \\
		\hline
		Inner radius of the dust shell $R_{in}/10^3~R_{\odot}$ & $1.7$ \\
		$R_{in}$ to stellar $R_{\star}$ ratio & $6.4$\\
		Dust temperature at shell outskirt $T_d$ (K) & 23  \\
		mass-loss rate (dust+gas) $\dot{M}/10^{-5}~M_{\odot}~{\rm yr}^{-1}$ & $1.3 \pm 0.4 $\\
		Dust mass-loss rate  $\dot{M}_{dust}/10^{-8}~M_{\odot}~{\rm yr}^{-1}$  & $3.9 \pm 0.3$\\
		Terminal wind velocity $v_\infty$ (km s$^{-1}$) & $18 \pm 5$\\
		Bolometric luminosity $\log (L_{bol}/L_\odot)$ & $3.9 \pm 0.1$\\
		Bolometric magnitude $M_{bol}$ (mag) & $-4.9 \pm 0.3$ \\
		\hline
	\end{tabular}
\end{table}

At this point, an extra infrared excess in the Figure \ref{sed}  SED, peaking at $\sim 10^{12}$ Hz (or about $60\, \mu$m), remains to be accounted for.
We attempted different multilayer models with DUSTY, but with physically unacceptable results. 
Alternatively, this far infrared feature could be explained (see dotted orange line) 
with an additional shell of cold dust (with temperature $T_{cd} = 35$ K, comparable to the value at the outer shell in Table \ref{tablaDUSTY}) amounting to a mass of  $M_{cd} = 10^{-2} M_\odot$ according to the simple modified blackbody model of \citet{1981Dwek} with emissivity index $\beta = 1$, similar to that observed in dense circumstellar envelopes \citep{1989Weientraub, 1991Beckwith, 1993Knapp}. 
This dust is probably of smaller size \citep{2003Draine}, and could come from previous episodes of mass-loss, located above the material responsible for radio emission, just on top of the shell modeled using DUSTY and giving rise to the near-infrared emission.

As the SED is dominated by the optical-infrared emission, the DUSTY output  parameters for \irc\ in Table \ref{tablaDUSTY} are expected to be reliable. 
For instance, the derived  terminal wind velocity is in close accordance with the value assumed to model the radio part of the SED.
In addition, the model output puts the inner radius of the circumstellar shell very close to the stellar surface, $R_{in}  \simeq 7 R_\star$, which is the expected behaviour for
cool, evolved stars, where  the dust condensates out of stellar wind where the temperature falls enough, near $R_\star$ \citep{vanLoon1}.

\subsection{Further insight  photometric considerations}

The bolometric luminosity $\log(L_\mathrm{bol}/L_\odot) = 3.9$  in Table \ref{tablaDUSTY} has been estimated from the SED of a dust-enshrouded star, but assuming that the absorbed optical radiation is re-emitted at longer wavelengths.
Therefore, this value must be 
consistent with 
the intrinsic \irc\ luminosity. In fact, the integration of the SED in Figure \ref{sed} yields the same result. 
This luminosity, and its corresponding $M_{bol} = -4.9$,  are too low for an RSG, so \irc\ is plausibly an AGB.
Our previous photometric analysis in Subsection \ref{clasi}  pointed in the same direction.

Although our source is clearly variable, we can also cautiously 
explore colour-colour and colour-magnitude diagrams of the interstellar corrected magnitudes to clarify its nature.
For \irc\ the near-infrared colours are $(J-K_s) = 3.78$,  $(J-H) = 2.16$, and $(H-K_s) = 1.62$ (and their corresponding values  in Bessell \& Brett system, $(J-K) = 3.90$,  $(J-H) = 2.21$, $(H-K) = 1.59$ mag  \citep{2001Carpenter}), while the absolute M$_K = -8.29$. These extremely reddened values suggest our source is heavily obscured by circumstellar matter, and can be classified\footnote{Some of the studies published in the literature are referred to Magallanic Clouds, so we correct for the distance modulus in each case for an adequate comparison.} as dust-enshrouded extreme AGB \citep[\textit{i.e.}][]{2001OLIVIER,vanLoon1,2011Boyer,2020GRO}.
More sophisticated photometric diagram has been recently suggested  using an extinction-free colour ($W_{RP,BP-RP}-W_{K_s,J-K_s}$) defined in \citet{202Abia}, which for \irc\ amounts to $3.76$, reinforcing the hyphotesis of dust-enshrouded AGB \citep{2018Lebzelter,2020GRO,2023Messineo}.

It is possible to determine the radius of \irc\ from the luminosity and the Stephan-Boltzmann law, yielding $R_\star = 233$ $R_\odot$, which is not far from the value derived from the DUSTY model (see Table \ref{tablaDUSTY}), and typical for an AGB star.

\subsection{Mass-loss and other estimates from DUSTY modeling and photometric data}

The mass-loss rate, which is a key parameter in AGBs and RSGs as it helps us in understanding the late stages of star evolution.
Our  combined DUSTY model fit (see Table \ref{tablaDUSTY})
renders $\dot{M} = (1.3\pm0.4) \times 10^{-5}M_\odot$ yr$^{-1}$, in excellent agreement with the value obtained above from the radio emission.
Moreover, we can estimate the mass-loss rate from photometric data using the prescription for dust-enshrouded AGBs of \citet{1989JURA}.
For the $60\, \mu$m flux of $65.54$ Jy, together with the assumed distance and wind velocity, we get again a very consistent $\dot{M} = 1.02\times 10^{-5} M_\odot$ yr$^{-1}$.
Here, we have used the bolometric luminosity in Table \ref{tablaDUSTY} and a mean wavelength in the SED in units of $10\, \mu$m, $\lambda_{10} \simeq 0.7$ as colour ($K-[12]$)$=4.1$, being $[12]$ the twelve micron magnitude  \citep{2001OLIVIER}. Again, all these rates are plausible for our scenario of \irc\ being a dust-enshrouded AGB.
These compatible values from different methods give us confidence on the results.

It is also interesting to derive the bolometric magnitude  from the period–luminosity relation for dust-enshrouded AGBs in our Galaxy \citep{2001OLIVIER, Karambelkar_2019}, 
assuming that \irc\ is a fundamental mode pulsator. The resulting M$_{bol} \simeq -5$ is not far from the obtained from the SED analysis, and compatible with other AGB period-luminosity relations in Magellanic Cloud \citep{2020GRO}.

Period or bolometric luminosity in LPVs could additionally be related to the actual stellar mass from different prescriptions \citep[\textit{i.e.}][]{1965Gough, 1983Wood, 2013Take}, rendering always $M_\star \sim 1 M_\odot$,  compatible with the suspected AGB origin.



\subsection{The nebula around \irc}

\begin{figure}
\epsscale{0.9}
	\plotone{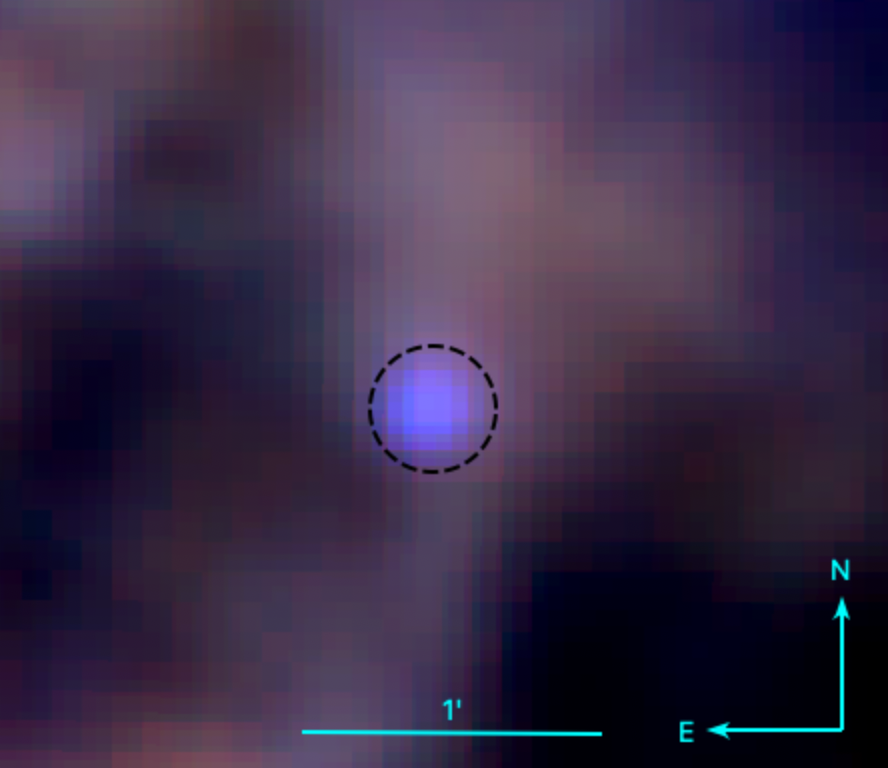}
	\caption{Trichromatic \irc\ field showing cold dust enshrouding the source. Blue is
		\textit{Herschel} PACS $70\, \mu$m emission, green is  \textit{Herschel} PACS $160\, \mu$m and red is \textit{Herschel}  SPIRE $250\, \mu$m.
		 Black circle marks the position of the nebula around \irc. 
		\label{3col}}
\end{figure}


Successive episodic outflows from old, evolved stars usually create spherical envelopes around the source.
Thus, thermal emission from dust in the circumstellar environments may be evidence of  mass-loss episodes.
For \irc\ we can observe the cold dust in the \textit{Herschel} PACS $70~\mu$m image (Figure \ref{3col}).
Inside the black circle we can find the compact dust shell enshrouding the source, with a mean angular radius of $\sim 9^{\prime\prime}$. 
This value should be similar to the outer radius of the shell, where the cold dust has a temperature of about  $T_d = 23$ K (see Table \ref{tablaDUSTY}). 
Given that we assumed $\xi= 10^4$, and that the distance to the source is $1.6$ kpc, the angular size agreement is remarkable.

\section{Conclusions.}

In this paper we have presented the discovery of \irc\ as an out of common LPV.
The reported photometry shows a variability amplitude not as high as to be considered a Mira, but 
clearly fitting the criteria to be considered a semiregular variable of SRb type.

The most prominent feature of \irc\, rendering it a singular case among LPVs, is its strong radio emission with $\nu^{+0.6}$ dependence.
Giants and supergiant stars radio fluxes are typically much weaker than the values
observed for \irc. This peculiarity makes this target stand out strongly among its peers and worth of additional follow-up work.
We speculate that the strong radio luminosity of \irc\ could be due to its powerful wind,
when completely ionized by the nearby OB associations (Serpens 2 and Scutum 3).
This allows us to estimate its mass-loss rate in a completely independent manner directly from radio emission.
Comparison with other traditional methodologies used
 for AGBs and RSGs, which involve empirical prescriptions or optical-infrared SED fitting has been satisfactory, thus adding reliability to out knowledge
 of this important parameter. 

The high optical depth, effective temperature, reddish colours, and other properties derived from our SED and photometric analysis of \irc\  make it to be plausibly a heavily absorbed, 
low-mass, dust-enshrouded AGB of spectral type M, as originally proposed by \citet{1985ApJS...58..167L}.
The period-luminosity relations also support the membership of this source to this scarce kind of cool, evolved stars. 
Its bolometric luminosity is $\log(L_\mathrm{bol}/L_\odot) = 3.9$, while its estimated mass-loss rate
using different methods consistently points to a result very close to  $10^{-5} M_\odot$ yr$^{-1}$ .






\begin{acknowledgments}
Authors acknowledge support from project 
PID2022-136828NB-C42 funded by the Spanish MCIN/AEI/ 10.13039/501100011033 and “ERDF A way of making Europe".
Partly based on observations obtained with the Samuel Oschin Telescope 48-inch and the 60-inch Telescope at the Palomar Observatory as part of the Zwicky Transient Facility project. ZTF is supported by the National Science Foundation under Grants No. AST-1440341 and AST-2034437 and a collaboration including current partners Caltech, IPAC, the Oskar Klein Center at Stockholm University, the University of Maryland, University of California, Berkeley , the University of Wisconsin at Milwaukee, University of Warwick, Ruhr University, Cornell University, Northwestern University and Drexel University. Operations are conducted by COO, IPAC, and UW.
This research has made use of the SIMBAD database,
operated at CDS, Strasbourg, France
This work has made use of data from the European Space Agency (ESA) mission
{\it Gaia} processed by the {\it Gaia} Data Processing and Analysis Consortium, 
the Wide-field Infrared Survey Explorer (WISE;  funded by UCLS and JPL/CIT (NASA)),
the  2MASS (funded by University of Massachusetts and IPAC/CIT (NASA)),
the SDSS (funded by the Alfred P. Sloan Foundation, the NSF, NASA, the Japanese Monbukagakusho, the Max Planck Society, and the Higher Education Funding Council for England),
the Midcourse Space Experiment (MSX; funded by the Ballistic Missile Defense Organization with additional support from NASA),
AKARI (a JAXA project with the participation of ESA),
and the Pan-STARRS1 Surveys (PS1, funded from the members of its consortium (\url{https://ippc20.ifa.hawaii.edu/pswww/?page_id=226})).
This research has also made use of the NASA/ IPAC Infrared Science Archive, which is operated by the Jet Propulsion Laboratory, California Institute of Technology, under contract with the National Aeronautics and Space Administration. This research has also made use of the VizieR catalogue access tool, CDS, Strasbourg, France.


\end{acknowledgments}

%









\bibliography{biblio3}{}
\bibliographystyle{aasjournal}



\end{document}